\documentclass[]{JHEP3} 

\preprint{}

\usepackage{epsfig,multicol}

\newcommand\fverb{\setbox\pippobox=\hbox\bgroup\verb}
\newcommand\fverbdo{\egroup\medskip\noindent%
			\fbox{\unhbox\pippobox}\ }
\newcommand\fverbit{\egroup\item[\fbox{\unhbox\pippobox}]}
\newbox\pippobox

\title{Indirect Detection of Dark Matter WIMPs in a Liquid Argon TPC}

\author{A. Bueno, R. Cid and S. Navas-Concha \\
	Dpto de F{\'\i}sica Te\'orica y del Cosmos \& C.A.F.P.E. \\ 
	Universidad de Granada, Granada, Spain \\
	E-mail: \email{a.bueno@ugr.es}, \email{rosalia@ugr.es}, \email{navas@ugr.es}}

\author{D. Hooper \\ 
	Astrophysics Department \\
	Univeristy of Oxford, OX1 3RH Oxford, UK \\
	E-mail: \email{hooper@astro.ox.ac.uk}}

\author{T. J. Weiler \\ 
	Department of Physics and Astronomy\\
	Vanderbilt University, Nashville TN 37235, USA \\
	E-mail: \email{t.weiler@vanderbilt.edu}}

\abstract{We assess the prospects for indirect detection
of Weakly Interacting Massive Particles using a large
Liquid Argon TPC detector. Signal events will consist
of energetic electron (anti)neutrinos coming from the decay
of $\tau$ leptons and $b$ quarks produced in WIMP annihilation in 
the core of the Sun. Background contamination from
atmospheric neutrinos is expected to be low, thanks
to the superb angular resolution and
particle identification capabilities provided
by the considered experimental set-up. We evaluate the event rates predicted for an annihilating WIMP as a function of its elastic scattering cross section with nucleons. This technique is particularly useful for WIMPs lighter than $\sim100$ GeV.}


\begin{document} 


\section{Introduction}
\label{sec:intro}

Astrophysical observations, probing gravitational potentials, 
provide overwhelming evidence for the
existence of matter that does not emit or absorb
electromagnetic radiation and therefore is known as {\it dark matter} (for recent
reviews see Refs.~\cite{Munoz:2003gx,Bertone:2004pz,Sumner:2002}). Perhaps the most convincing evidence for the existence of dark matter comes from the rotation
curves of a large number of spiral
galaxies~\cite{Persic:1995ru,SOFUE:2000jx}. Additional data, such as that collected by WMAP~\cite{Spergel:2003cb}, favours a standard cosmological
model in which the Universe is flat and contains about 25$\%$
non-baryonic dark matter~\cite{Freedman:2003ys}.

Notwithstanding the large set of independent observations favouring 
its existence, the exact nature
of dark matter remains one of the
most puzzling mysteries of our time. Particle Physics, through
extensions of the Standard Model, can provide potentially viable candidates for dark
matter, most of which fall in the class of Weakly Interacting Massive Particles
(WIMPs). WIMPs are stable (or very long-lived)
particles, with masses ranging from the GeV to the TeV range, which can account for a
significant fraction of the dark matter density in the Universe. Among
the plethora of WIMP candidates, the lightest neutralino is the most
theoretically
developed~\cite{Goldberg:1983nd,Krauss:1983ik,Ellis:1983ew,Gondolo:2004sc}.
Another possibility is offered by Universal Extra Dimension
models~\cite{Appelquist:2000nn}, in which the Lightest Kaluza-Klein particle
is a plausible dark matter candidate~\cite{Servant:2002aq}. 
  
Parallel to theoretical developments, there also exists an intense
and challenging experimental activity devoted to WIMP detection. So
far, only a single experiment, DAMA~\cite{Bernabei:2000qi}, claims 
to have found evidence for the presence of WIMPs in the
Galactic halo. Using a direct detection method, they have measured an 
annual modulation, over seven annual cycles (107,731 kg day total
exposure), consistent with expectations from a WIMP signature. Other
experiments, probing similar regions of the parameter space, have
found negative results~\cite{Akerib:2004fq,Chardin:2003vn,zeplin,Ahmed:2003su}. Therefore the situation remains highly controversial. Indirect
detection methods have produced so far negative
results~\cite{Ambrosio:1998qj,Desai:2004pq}.
 
In this paper we concentrate on indirect detection methods. 
We assess, in a model-independent way, the capabilities that
a massive Liquid Argon (LAr) Time Projection Chamber (TPC) offers for identifying
neutrino signatures coming from the products of WIMP annihilations in the core of the Sun. 
Unlike existing measurements
and future searches in large \v{C}erenkov neutrino
telescopes~\cite{icecube,Halzen:1998sc,Blanc:2003na,Bottai:1998gh}, 
where a statistically significant excess of high energy
muons is expected to be measured, here we propose to 
look for an excess of electron-like events over the
expected background of cosmic neutrinos. In particular, 
our signature will be a high energy $\nu_e$ and $\bar\nu_e$
charged current interactions pointing in the direction of the Sun. 
We evaluate our sensitivity and present the expected event rates over the mass region of 10 to 100~GeV where this technique is most effective. Note that this includes the mass range favoured by
the DAMA experiment~\cite{Bernabei:2000qi}.


\section{Neutrinos from WIMP Annihilation in the Sun}
\label{sec:WIMPsun}

WIMPs that constitute the halo of the Milky Way can occasionally interact with massive objects, such as stars or planets. When they scatter off of such an object, they can potentially lose enough energy that they become gravitationally bound and eventually will settle in the center of the celestial body. In
particular, WIMPs can be captured by and accumulate in the core of the Sun. 

The capture of heavy particles in the Sun was first calculated
in Ref.~\cite{Press:1985ug}. Improved formulae for WIMP capture were later given in Ref.~\cite{Gould:1987ir}. WIMPs can accrete as a result of both spin-dependent and spin-independent (scalar) interactions. The solar WIMP capture rate is given by~\cite{Gould:1987ir}: 
\begin{equation}
C^\odot \simeq 3.35 \times 10^{20} \mathrm{s}^{-1}
\left(\frac{\rho_{\rm{local}}}{0.3 \ \mathrm{GeV/cm}^3}\right)
\left(\frac{270 \ \mathrm{km/s}}{\bar{v}_{\rm{local}}}\right)^3
\left(\frac{\sigma_{\rm{H,SD}}+\sigma_{\rm{H,SI}}+0.07\sigma_{\rm{He,SI}}}{10^{-6}\mathrm{pb}}\right)
\left(\frac{100 \ \mathrm{GeV}}{m_{\rm{WIMP}}}\right)^2
\label{eq:rate}
\end{equation}
where $\rho_{\rm{local}}$ is the local dark matter density,
$\bar{v}_{\rm{local}}$ is the local rms velocity of the halo WIMPs,
$\sigma_{\rm{H,SD}}$, $\sigma_{\rm{H,SI}}$ and $\sigma_{\rm{He,SI}}$ are the spin-dependent and spin-independent WIMP-Hydrogen and WIMP-Helium cross sections, respectively, and $m_{\rm{WIMP}}$ the mass of the WIMP. Throughout this paper, we will refer to the quantity $\sigma_{\rm{H,SD}}+\sigma_{\rm{H,SI}}+0.07\sigma_{\rm{He,SI}}$ simply as the elastic scattering cross section, $\sigma_{\rm{elastic}}$.


In this paper, we do not consider the possibility of observing neutrinos from WIMPs accumulated in the Earth. Given the smaller mass of the Earth and the fact that only scalar interactions contribute, the capture rates for our planet are not
enough to produce, in our experimental set-up, a statistically
significant signal. For a detailed discussion of WIMP capture in the Earth, see Ref.~\cite{edsjoearth}.

Accumulated WIMPs will annihilate into ordinary matter (quarks,
leptons, gauge and Higgs bosons). In the case of neutralino dark matter, annihilations do not directly produce neutrinos, as they are Majorana particles~\cite{Goldberg:1983nd}. Instead, neutrinos with a spectrum of energies below the mass of the WIMP will be generated in the decays of annihilation products. 

Other dark matter candidates may produce neutrinos directly, however. For example, Kaluza-Klein dark matter~\cite{Cheng:2002ej,Hooper:2002gs} or sneutrino dark matter \cite{sneutrinos,sneutrinos2} can each annihilate directly to neutrino pairs.


As already pointed out, we are only interested in the expected fluxes of
electron neutrinos and anti-neutrinos. To compute this, we use the
analytic expressions given in Ref.~\cite{Jungman:1994jr}. The
differential energy flux is: 
\begin{equation}
\frac{d\phi}{dE} = \frac{\Gamma_A}{4\pi R^2} \sum_F B_F \left(\frac{dN}{dE}\right)_F
\label{eq:flux}
\end{equation}
where $\Gamma_A$ is the WIMP annihilation rate in the Sun and R is the
distance from the Sun to the Earth. The sum is over annihilation channels, $B_F$ being the fraction of annihilations which go to channel $F$. $(dN/dE)_F$ is the differential energy spectrum of electron neutrinos at the
surface of the Sun per WIMP annihilating through channel $F$.


The final electron (anti)neutrino flux is obtained using the following set of
assumptions: 
\begin{itemize}
\item We take typical values for the local dark matter density and
local rms velocity, namely $\rho_{local}=0.3 \ \mathrm{GeV/cm}^3$ and
$\bar{v}_{local}=270$ km/s.
\item Equilibrium has been reached and therefore the annihilation
rate, $\Gamma_A$, is equal to the capture rate in the Sun,
$C^\odot$. This is a reasonable assumption given the range of cross sections we consider here.
\item We restrict ourselves to a low mass WIMP scenario,
namely $m_{\rm{WIMP}} \leq 100$ GeV; in this way we minimize
neutrino flux depletion due to propagation effects across the Sun.

\item Due to the former assumption, phase space considerations avoid
WIMP annihilation into top quarks and Higgs bosons. Although in the narrow mass range between 80 and 100 GeV, annihilations to gauge boson pairs ($W^+W^-$, $Z^0 Z^0$) are possible, we do not explore this channel. We instead focus on WIMP annihilations to the $b \bar{b}$ and $\tau^+ \tau^-$ channels. These are generally the dominant channels for a neutralino in this mass range. 

The electron neutrino spectrum from annihilations to $\tau^+ \tau^-$ is given (in the relativistic limit) by~\cite{Jungman:1994jr}
\begin{equation}
\left(\frac{dN}{dE}\right)_{\tau^+\tau^-} = \left(\frac{2\Gamma_{\tau\to
e\nu\bar{\nu}}}{m_{\rm{WIMP}}}\right) (1-3x^2+2x^3)\,\, e^{-E/E_k},
\label{eq:taurate}
\end{equation}
where $x=E/m_{WIMP}$ and E is the neutrino energy. $\Gamma_{\tau\to
e\nu\bar{\nu}} \simeq 0.18$, the branching ratio of taus into an electron plus neutrinos. The factor, $e^{-E/E_k}$, accounts for the absorption of neutrinos as they propagate through the Sun. $E_k \simeq 130$ and $200$ GeV for $\nu_e$ and $\bar{\nu_e}$, respectively~\cite{Albuquerque:2000rk,Crotty:2002mv}.

The electron neutrino spectrum from annihilations to $b \bar{b}$ is similar~\cite{Jungman:1994jr}
\begin{equation}
\left(\frac{dN}{dE}\right)_{b \bar{b}} = \left(\frac{2\Gamma_{b\to
e \nu X}}{z \, m_{\rm{WIMP}}}\right) (1-3x^2+2x^3)\,\, e^{-E/E_k}
\label{eq:brate}
\end{equation}
Here, $x=E/(z\, m_{\rm{WIMP}})$ where $z\cong 0.7$ is the fraction of energy retained after hadronization. The inclusive branching ratio for $b$ quarks to electron neutrinos, $\Gamma_{b\to e \nu X}$, is approximately 0.11.

\item  We consider that tau pairs are produced with a branching
fraction $B_{\tau^+ \tau^-}=10\%$ while for $b$ quarks we assume a branching ratio of $B_{b \bar{b}}=90\%$.
\end{itemize}


\section{The Liquid Argon TPC}
\label{sec:LAr}

The detection device we propose to use for indirect WIMP detection
is a Liquid Argon Time Projection Chamber (LAr TPC). This technology 
was first put forward in 1977~\cite{crub}. In a LAr TPC, the ionization
charge, released at the passage of charged particles, can be transported over 
large distances thanks to the presence of a uniform electric
field. The signals induced by drift electrons are recorded by a set of
successive anode wires planes thus providing three dimensional image 
reconstruction and calorimetric measurement of
ionizing events. Unlike traditional bubble chambers,
limited by a short window of sensitivity
after expansion, the LAr TPC detector remains fully and continuously 
sensitive, self-triggerable and without read-out dead time.

The feasibility of the LAr TPC technology has been demonstrated
thanks to an extensive R\&D programme, developed by the ICARUS
collaboration, that culminated in the construction and operation of a
600 ton detector~\cite{Amerio:2004}. A surface test of this detector
has demonstrated the high level of maturity reached by this
technology~\cite{Amoruso:2004dy,Amoruso:2004ti,Antonello:2004sx,Amoruso:2003sw,Arneodo:2003rr}.

Future applications of this detector technology have been proposed to
study neutrino properties within the context of neutrino
factories~\cite{Bueno:2000fg,Bueno:2001jd} or to study neutrinos from 
supernovae~\cite{Gil-Botella:2003sz}. An interesting long-term
application is the possibility to build a giant LAr TPC detector with
a mass of 100 ktons. The technical aspects and physics prospects of
such detector have been extensively discussed
elsewhere~\cite{andre:2003,Ereditato:2004ru}. In the present paper we
evaluate for the first time the capabilities that such a detector will have
in the search for dark matter WIMPs.

We have focused our interest on $\nu_e$, $\bar{\nu}_e$ charged current events, 
since a LAr TPC offers very good electron identification capabilities. In
addition, all final state particles can be measured and identified
(see figures~\ref{fig:event_coll} and~\ref{fig:event_ind} where a full
simulation of a $\nu_e$ charged current interaction is shown). Thanks to
the high granularity and superb imaging capabilities, we expect to
have a very accurate measurement of both energy and incoming
direction. Therefore our signal will
correspond to a neutrino interaction that points in the direction of
the Sun and contains an energetic primary electron. Possible
background sources are extensively discussed in the next section.

 \FIGURE{\epsfig{file=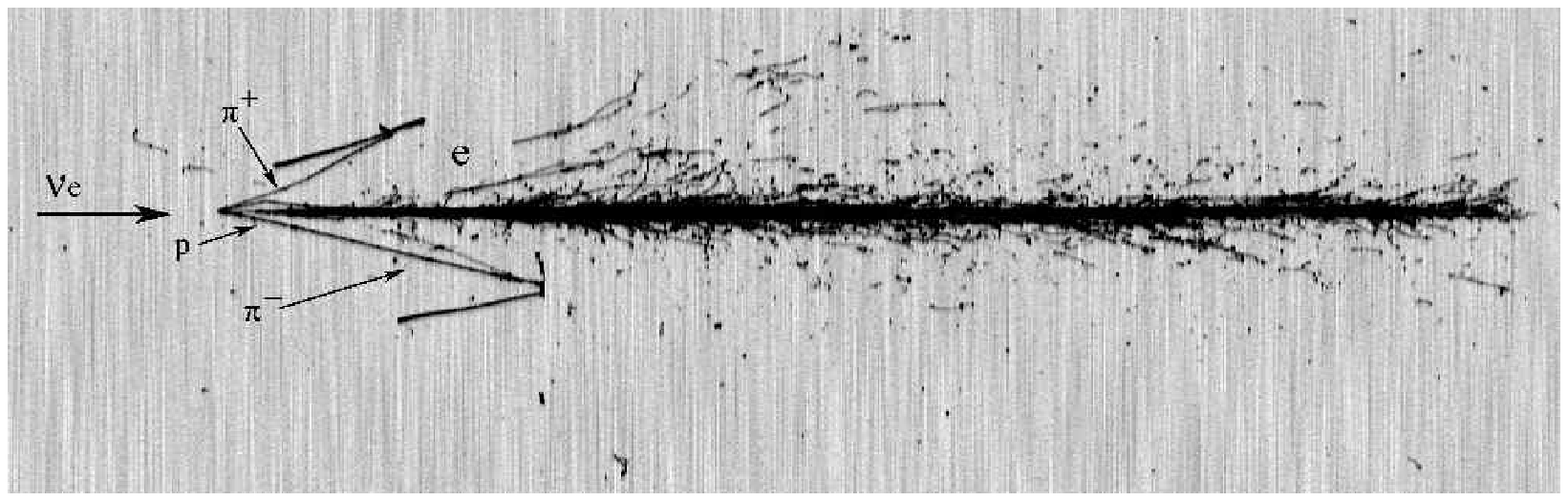,
         width=13cm,keepaspectratio} 
 \caption{Simulated $\nu_e$ interaction in a Liquid Argon detector
  (longitudinal view). The electron neutrino ($E_\nu$ = 26.3~GeV)
  travels from left to right and suffers a charge current interaction
  with the atoms of the medium, resulting in a highly energetic electron
  (23~GeV), two pions ($E_{kin} =$~0.7 and 1.4~GeV) and a proton
  ($E_{kin} =$~1.6~GeV). This last particle is not visible in this projection.
  The neutrino direction is computed combining the information
  coming from the electromagnetic shower and the hadronic jet.}
 \label{fig:event_coll}}
 
 \FIGURE{\epsfig{file=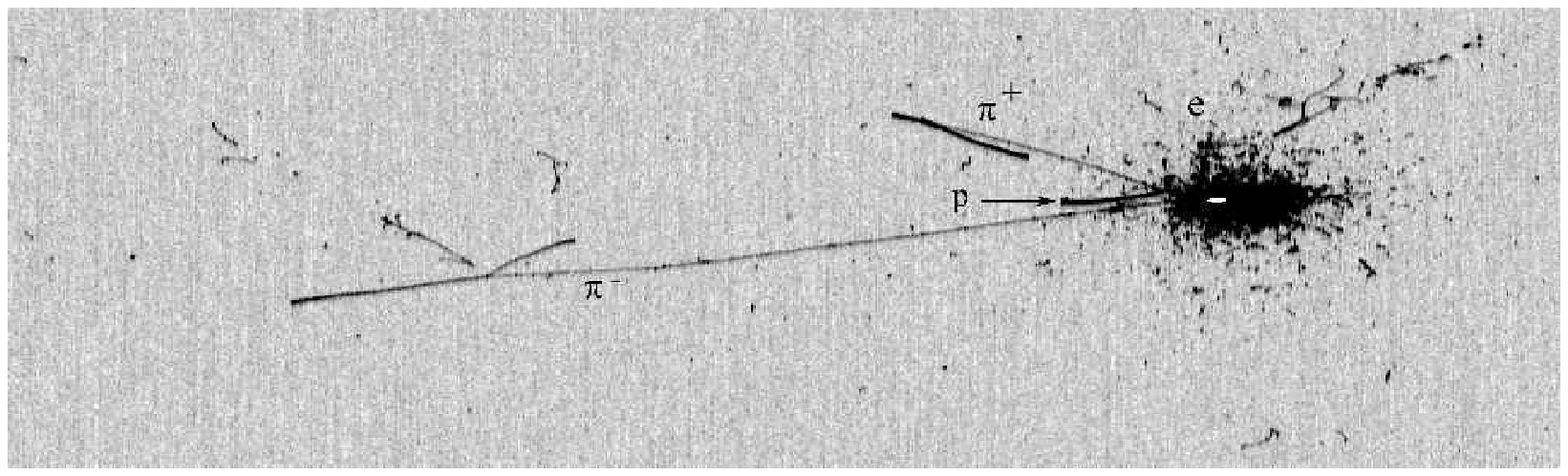,
         width=13cm,keepaspectratio} 
 \caption{Rotated view of the same $\nu_e$ interaction shown
  in figure~\ref{fig:event_coll}. In this projection the four particles
  coming out from the main interaction vertex are clearly visible:
  two pions, one proton and the electron (that appears as a big black spot).}
 \label{fig:event_ind}}

\section{Sources of $\nu_e$ Background}
\label{sec:background}

We have identified several sources that could, a priori, significantly
contribute to the expected background rate. The most
obvious is atmospheric neutrinos, where we expect
contributions from both $\nu_e / \bar{\nu}_e$ charged currents and 
$\nu / \bar{\nu}$ neutral currents (where a $\pi^0$ is 
misidentified as an electron). In the energy range we study here ($\sim$10 to 100~GeV) solar neutrinos do not significantly contribute to the  background. We have also considered the flux of
neutrinos originating from cosmic ray interactions with the
interstellar medium in the galaxy. As shown in Ref.~\cite{Ingelman:1996md},
for the range of energies considered in this study, this flux is 
negligible compared to the atmospheric flux. The diffuse flux of
$\nu_e$ from active galactic nuclei is also small. 

Fluxes of $\nu_\tau$ are also a concern given that $\tau$ leptons, 
produced in charged current interactions, can subsequently decay into
electrons. According to Ref.~\cite{Athar:2004uk}, the expected flux of
$\nu_\tau$ from sources like atmospheric neutrino oscillations, the galactic plane
and galaxy clusters are small and will not give a significant
contribution in our energy range. 

Therefore, the most important source of background is by far due
to atmospheric neutrinos. The evaluation of this flux $\phi_\nu$ has been
performed using the calculations published in Ref.~\cite{battistoni}.
In this paper the authors present a detailed 3-dimensional
calculation of the atmospheric neutrino flux based on the
{\tt FLUKA} Monte Carlo model~\cite{FLUKA}.
The final {\tt FLUKA} tables for three experimental sites
(Kamiokande, Gran Sasso and Soudan) are also accessible
in Ref.~\cite{battistoni:web}.

 Figure~\ref{fig:fondo_vs_energy_angle} shows the differential atmospheric
electron neutrino flux as a function of the neutrino zenith angle and the neutrino
energy expected at the latitude of Gran Sasso~\cite{battistoni:web}.
Similar distributions are built and used for $\bar{\nu}_e$, $\nu_\mu$ and
$\bar{\nu}_\mu$. Only neutrinos in the energy range 1~--~100~GeV are shown.

 Without loss of generality, the spectra at Gran Sasso has been taken
as reference for all calculations presented in this work. The conclusions of
the paper remain valid for other underground laboratories
such as Kamiokande or Soudan since the neutrino fluxes at these locations
are expected to be similar (at high energies, differences are less than
a percent~\cite{battistoni:web}).


\FIGURE{\epsfig{file=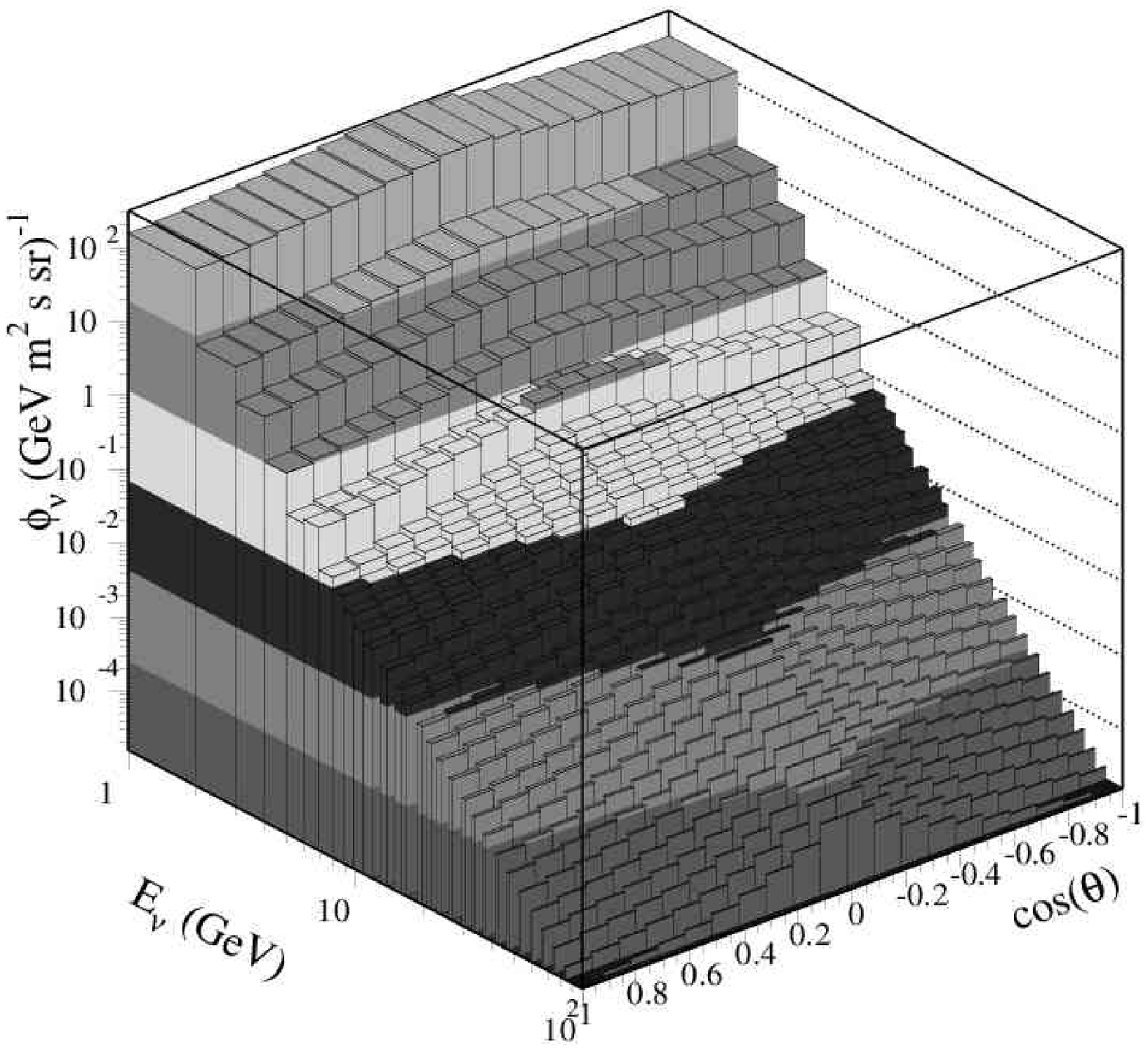,
        width=12cm,keepaspectratio}
\caption{Differential atmospheric electron neutrino flux as a function of the
 neutrino energy and zenith angle.
 The data, taken from Refs.~\cite{battistoni,battistoni:web}, correspond
 to the Gran Sasso latitude and is taken as our reference to 
 compute the expected background rate.}
\label{fig:fondo_vs_energy_angle}}


\section{Data Analysis}
\label{sec:analysis}

 The number of expected background events contributing from each neutrino
species ($N_{\nu_i}$) is determined from simple integration of the associated
atmospheric differential neutrino flux ($\phi_{\nu_i}$) taking into account
the involved cross sections ($\sigma_{\nu_i}$),
the detector mass ($M_d$) and the data taking period:

\begin{equation}
N_{\nu_i} = \int \phi_{\nu_i} (E,\theta) \cdot
    \sigma_{\nu_i}(E) \cdot M_d \;\; dE \; d\Omega \; dt,
 \label{eq:integral}
\end{equation}

\noindent
where the integration is done above a minimum energy threshold,
$E^{min}_{\nu}$, and inside a limited solid angle on the sky, $\Delta\Omega$,
in order to reduce the background contribution.
In the following we assume a detector mass of 100~ktons of LAr and 10 years
of data taking. 

We have taken into account
not only Charged Current (CC) events for $\nu_e$ and ${\bar{\nu}}_e$ interactions 
but also Neutral Current (NC) events for both electron and muon species.
Neutral current events contribute to the background mainly through (1) electrons
from Dalitz decays and (2) misidentified pions. In this work we have assumed
a conservative 90\% efficiency on the rejection of NC events,
even though Monte Carlo studies on LAr detectors show much better capabilities
($>$99\%).
$\nu_\mu$ and ${\bar{\nu}}_\mu$ CC interactions are
excluded from the calculation since the presence of a muon in the final state
would immediately veto the event as a possible $\nu_e$ 
(the muon--electron misidentification probability in a LAr detector
is almost zero). 

 In general, when looking at neutrinos coming from astrophysical point--like
sources, the atmospheric neutrino background can be reduced to a very low 
level by selecting small angular regions
of the sky. This applies to our particular problem where we are interested
in neutrinos coming only from the Sun.
Therefore the region of interest is of the order of the angular size of the Sun and, in principle, the neutrino background coming from this narrow sky window
will be small.

 At this stage we have to estimate $\Delta\Omega$, the apparent size
of the Sun ``seen'' by the detector. Its value will be driven by our
capabilities to determine the incoming neutrino direction, i.e.
the detector angular resolution.
This calculation has been carried out through a detailed
and realistic simulation of neutrino interactions inside a LAr detector.
Neutrinos of different species in a wide energy range ($< 100$~GeV)
and the final state particles, produced after the $\nu$ interaction,
were tracked using the {\tt FLUKA} package.
 As an example, figures~\ref{fig:event_coll} and~\ref{fig:event_ind}
show a picture of one of these simulated events where a 26.3~GeV
electron neutrino (flying from left to right)
undergoes a CC interaction giving rise to several secondary particles:
$\nu_e \; n \rightarrow e^- \; p \; \pi^+ \; \pi^- \; (2n \, 3 \gamma)$.
The three charged hadrons appear as dark thin lines over light grey background,
the electron develops as an electromagnetic shower and the neutral particles
are invisible. The reconstruction of the $\nu$ direction is done using the
information coming from all particles produced in the final state.

 The result of the simulation is shown in figure~\ref{fig:ang_resol},
where the difference between the simulated and reconstructed neutrino directions, $\Delta\theta$, 
is plotted as a function of the incoming neutrino energy
(solid curve). Note that for the considered energy range, O(10~GeV),
the smearing introduced by the Fermi motion of the initial state
nucleon and by re-interaction of hadrons inside the nucleus are negligible
and the improvement on the neutrino direction measurement achieved by combining
the electron and the hadronic jet informations is significant.
Due to the excellent reconstruction capabilities
offered by the LAr technique, the angular resolution remains below $2^\circ$
for the highest energies up to about 15~GeV, increasing by a factor of 4
at lower energies ($\sim$~5~GeV).

 In case of an ideal detector with perfect angular resolution,
the sky integration window would be limited to the size of the Sun
($\sim 6.7\cdot10^{-5}$~sr). In our more realistic scenario $\Delta\Omega$
has been estimated by smearing this number with the detector resolution.
The dashed line in figure~\ref{fig:ang_resol} shows the resulting
half size of the Sun (radius) after the smearing (see right axis).
For comparison, the real Sun radius seen from the Earth ($\sim 0.256^\circ$)
is also plotted. 

 Concerning the signal, the number of expected events is computed in a model
independent way as described in section~\ref{sec:WIMPsun} by integrating
equation~\ref{eq:flux} between $E^{min}_{\nu}$ and the WIMP mass.
Contributions from electron neutrino oscillations in the range of considered
neutrino energies (above 10~GeV) are negligible. Therefore, no loss or gain
on signal events from oscillated neutrinos is taken into account.  

 The final estimation for a 50~GeV WIMP mass is presented in
figure~\ref{fig:fondo_senal} where we show the number of expected signal (squares)
and background (dots) events as a function of the cut on the minimum neutrino energy.
The numbers are normalized to a 100~kton LAr detector and 10~years of data taking
and for a WIMP elastic scattering cross section in the Sun of $10^{-4}$~pb
(the result can be easily rescaled to other exposures or cross sections).
As expected, the background rate quickly decreases when increasing the neutrino
energy cut because of (1) the nature of the atmospheric neutrino flux
and (2) the better detector angular resolution.
From this figure we conclude that the search can be safely considered as
``background free'' for values of $E^{\rm{min}}_{\nu}$ above $\sim$~10~GeV
(contamination $\sim$~0.2~events).

 Figure~\ref{fig:f_s_3mass} gives the evolution of the signal events
as a function of the WIMP elastic scattering cross section, for
neutrinos of energy above 10~GeV and three values of the WIMP mass.
The background (horizontal line) is below the signal for values of
$\sigma_{\rm{elastic}}$ above $\sim 5\cdot10^{-6}$~pb.

\FIGURE{\epsfig{file=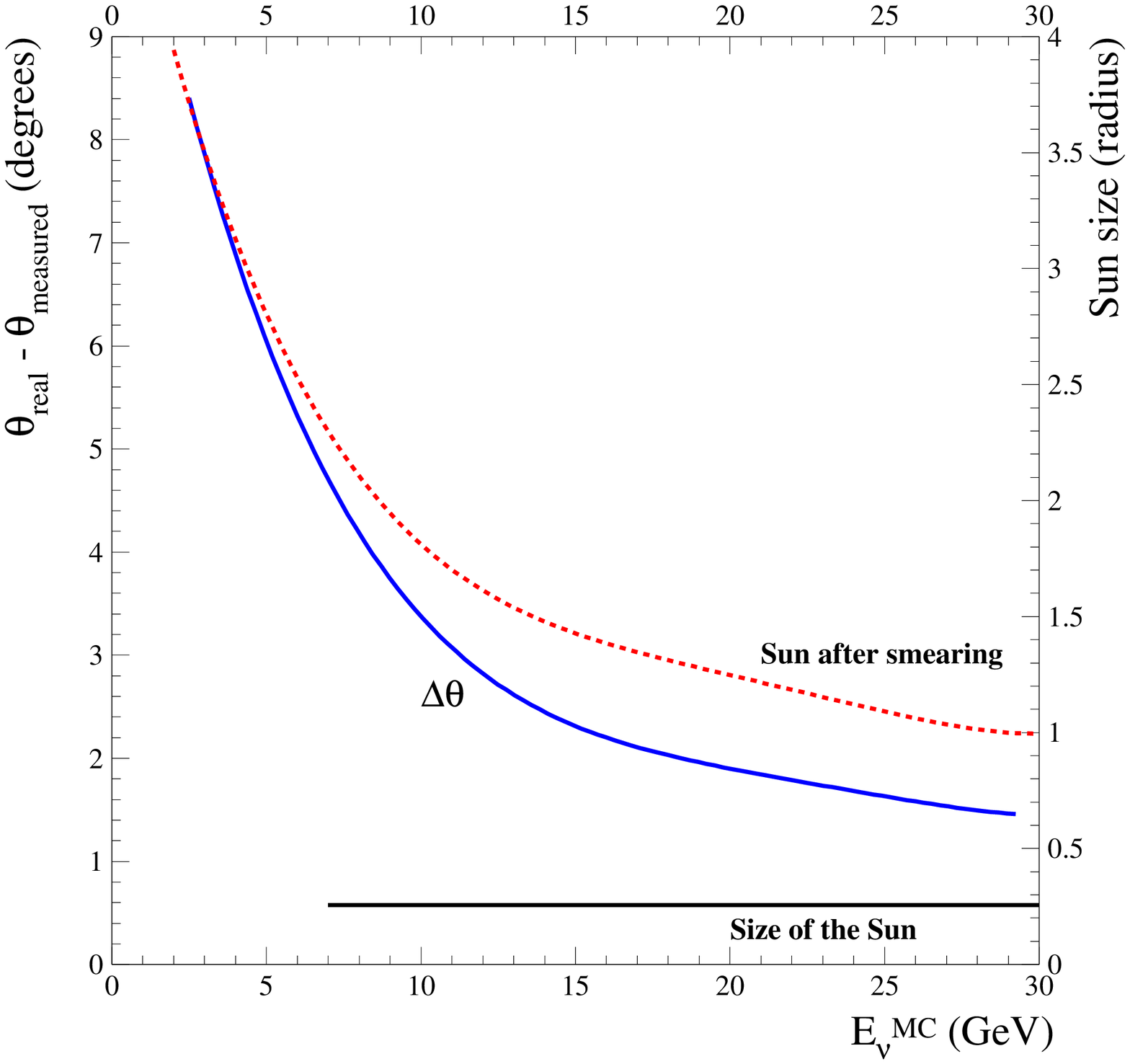,
        width=12cm,keepaspectratio} 
\caption{Detector angular resolution ($\Delta \theta$) as function
 of the incident neutrino energy obtained from a Monte Carlo
 simulation in a LAr detector (solid line and left axis).
 The figure also shows the size of the Sun (radius) for an ideal detector
 (horizontal solid line) and the resulting apparent size after detector
 smearing (dashed line, read on the right axis).}
\label{fig:ang_resol}}

\FIGURE{\epsfig{file=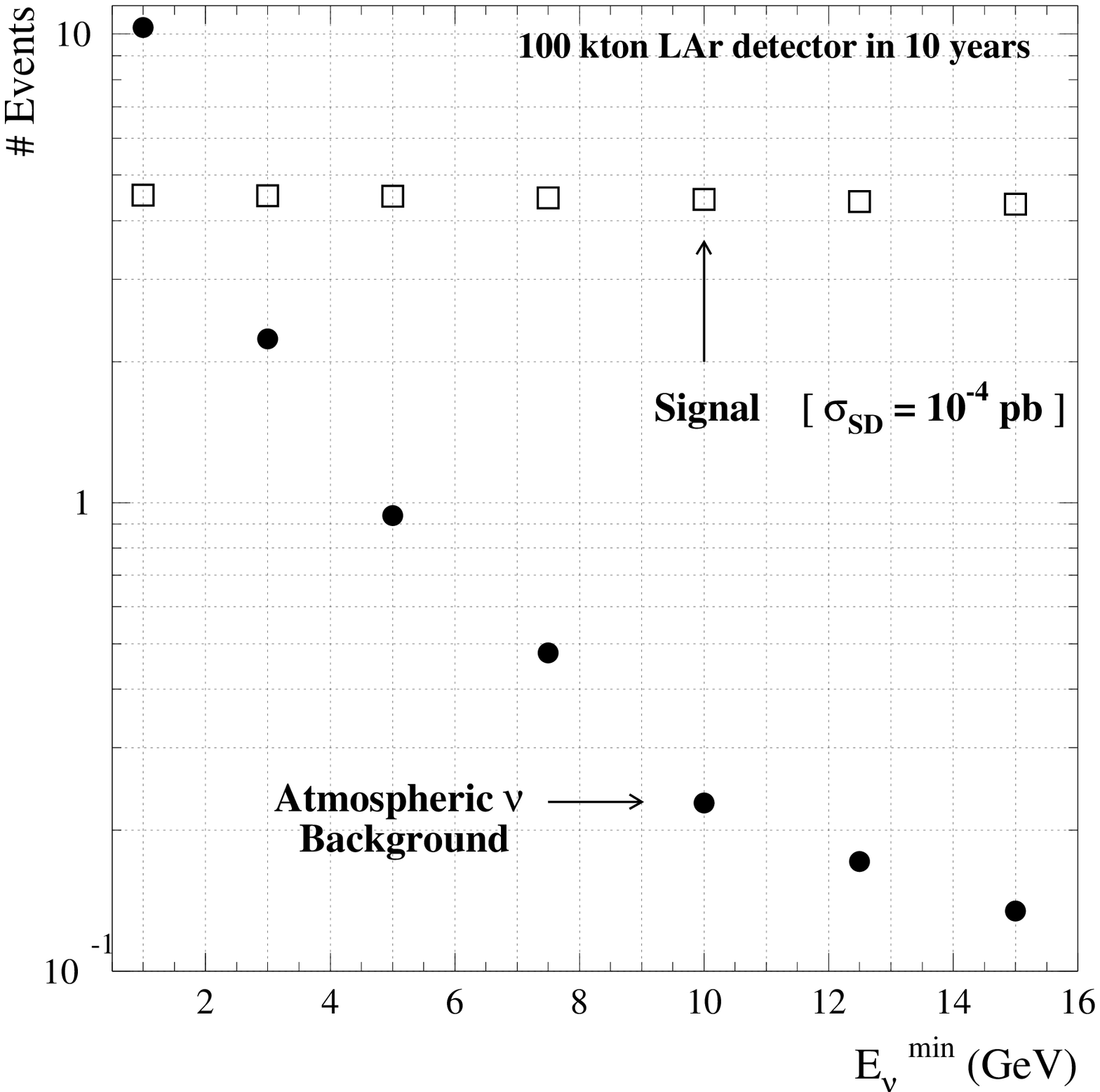,
        width=12cm,keepaspectratio} 
\caption{Expected number of signal and background events
 in a 100 kton LAr detector after 10 years of data taking
 as function of the cut on the minimum  neutrino energy.
 Signal events correspond to a 50~GeV WIMP mass and are normalized to a
 elastic scattering production cross section in the Sun of $10^{-4}$~pb.
 The number of atmospheric neutrino events includes
 contributions from muon and electron (anti)neutrinos, and takes into account
 the detector angular resolution.}
\label{fig:fondo_senal}}

\FIGURE{\epsfig{file=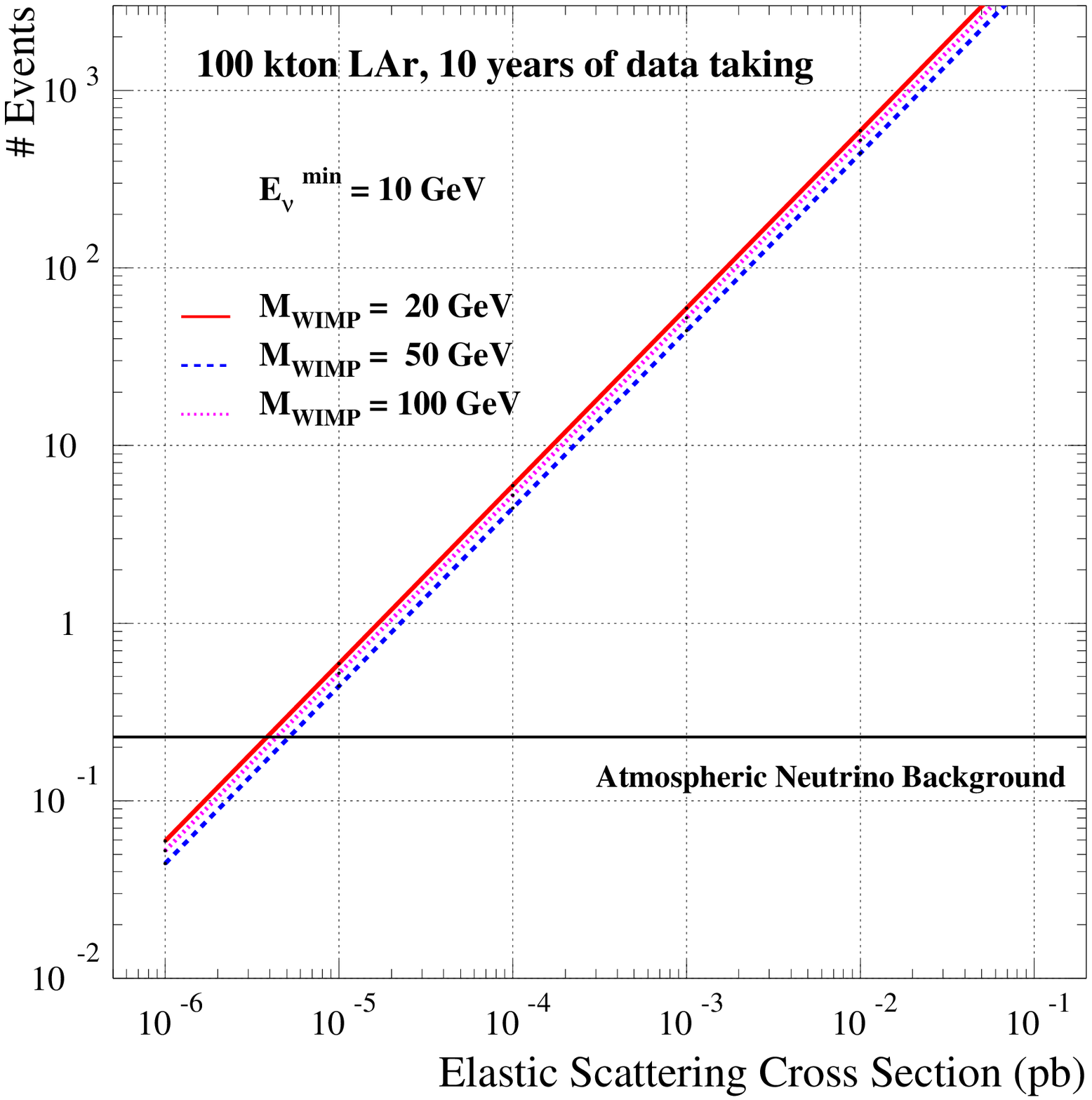,
        width=14cm,keepaspectratio} 
\caption{Expected number of signal and background events as a function of the
 WIMP elastic scattering production cross section in the Sun. The three lines correspond
 to three values of the WIMP mass and a cut on the minimum neutrino energy of 10~GeV.}
\label{fig:f_s_3mass}}

\section{Prospects for Discovery}
\label{sec:prospects}

 In section~\ref{sec:analysis} we have evaluated the 
average number of signal and background events 
(called hereafter $\mu_S$ and $\mu_B$, respectively)
expected in the detector for a certain exposure,
{\Large $\varepsilon$} = Mass $\times$ time = 100 ktons $\times$ 10 years.
At this point we would like to know whether a numbers
of expected events from a particular dark matter model will allow the experiment to claim a
discovery at a certain significance level.
This problem is fully discussed in Ref.~\cite{Hernandez:1996su},
where a general and well defined \emph{criterion of discovery} is
proposed for statistical studies of this nature:

\begin{itemize}
\item[-] Usually a 5$\sigma$ criterion (called ``{\tt method A}'' hereafter)
is used, so that in order to consider that a discovery will take place,
the signal must be above a 5$\sigma$ fluctuation of the background:
$\mu_S > 5 \sqrt{\mu_B}$.

 This commonly used rule frequently gives misleading (if not clearly wrong)
results. Only when $\mu_S$ and $\mu_B$ are large enough to describe
with Gaussian distributions can this rule be applied.   
This is clearly not the case in our study.

\item[-] Instead, we have followed the method
proposed in Ref.~\cite{Hernandez:1996su} (``{\tt method B}'' hereafter),
where the discovery criterion is given by two numbers, namely: $\epsilon$, the significance required to consider an excess of events a discovery, and $\delta$, the probability that the experiment will obtain such an excess.
For instance, when $\delta = 0.5$ and $\epsilon = 5.733\times10^{-7}$,
we can state that at least $50$\% of the time, the experiment will
observe a number of events which is $5\sigma$ or more above the background.


\end{itemize}

 The results based on ``{\tt method B}'' ($\delta = 0.5$ and
$\epsilon = 5.733\times10^{-7}$) are presented in
figure~\ref{fig:discovery_limit}.
 The plot shows the minimum number of years required
to observe a WIMP positive discovery signal as function of the WIMP elastic scattering cross section.
The numbers are given for three values of the WIMP mass and 
only neutrinos above 10~GeV are considered.

 From this result we conclude that, in 10 years of data taking,
a 100 kton LAr detector would be able to claim a clear discovery
signal according to ``{\tt method B}'', provided $\sigma_{\rm{elastic}}$
is above $\sim 10^{-4}$~pb for WIMP masses greater than 20~GeV.
On the other hand, only one year of data taking would be enough
to observe a discovery signal provided $\sigma_{\rm{elastic}}$ is above
$5\times10^{-4}$~pb and $m_{\rm{WIMP}} \sim$20~GeV.


Finally, we have studied the sensitivity of the results with the cut on the
minimum neutrino energy. We remind the reader that the value of $E_{\nu}^{\rm{min}}$
sets the energy above which all events are integrated. By lowering
the cut we would accept too many atmospheric background events and by
increasing it both, $\mu_S$ and $\mu_B$ decrease, but at different rates (see figure~\ref{fig:fondo_senal}). The quantitative result in terms of
discovery prospects are shown in figure~\ref{fig:discovery_limit_5-10gev_50cl}
for two values of $E_{\nu}^{\rm{min}}$ and a reference WIMP mass of 50~GeV.

\FIGURE{\epsfig{file=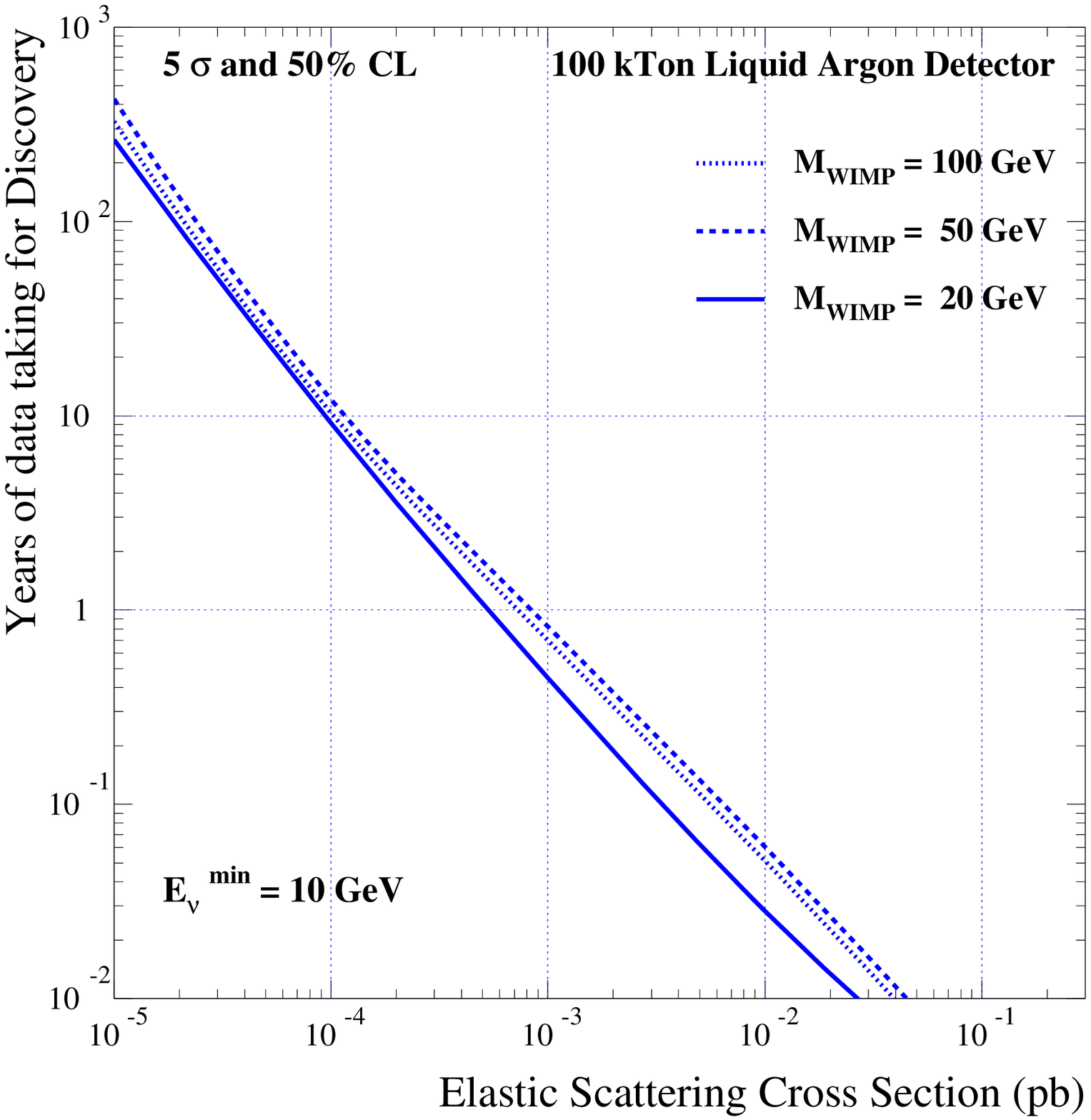,
        width=14cm,keepaspectratio} 
\caption{Discovery potential: 
 Minimum number of years required to claim a discovery WIMP signal
 from the Sun in a 100~kton LAr detector as function of $\sigma_{\rm{elastic}}$
 for three values of the WIMP mass. See text for details on the applied
 statistical criteria.}
\label{fig:discovery_limit}}

\FIGURE{\epsfig{file=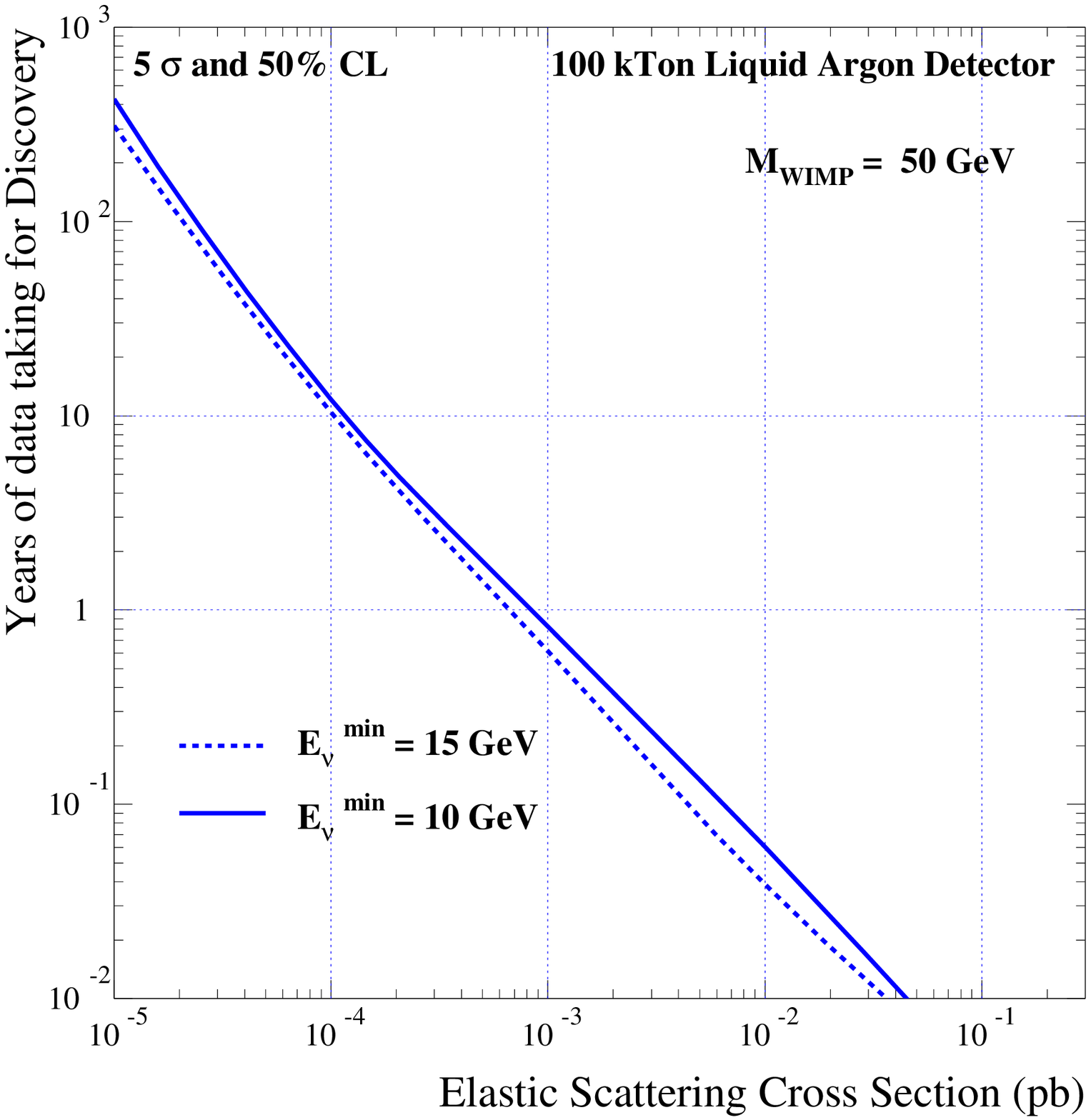,
        width=14cm,keepaspectratio} 
\caption{Detector discovery potential for two values of the minimum
 neutrino energy cut and a 50~GeV WIMP mass.
 See text for details on the applied statistical criteria.}
\label{fig:discovery_limit_5-10gev_50cl}}

\section{Comparison to Other Indirect Detection Techniques}
\label{sec:comparison}
Using neutrino detectors to search for evidence of dark matter has been studied 
in the context of several experiments. Currently, high energy neutrino
telescopes that use large volumes of a natural Cerenkov medium (such
as ice or water) show the greatest promise for dark matter detection. 
The AMANDA experiment, located at the South Pole, has been taking data
for several years. Its successor, IceCube is currently under
construction at the same site. When completed, IceCube, will constitute
the largest neutrino detector to date, with a full cubic kilometer of 
instrumented volume. The ANTARES experiment is currently under
construction in the Mediterranean. Although its effective volume will
be considerably smaller than IceCube's, ANTARES has been designed to
be sensitive to neutrinos of lower energy, and therefore may be 
competitive with IceCube for detecting dark matter.

At the energies we are considering here (well below a few TeV),
neutrino telescopes observe neutrinos through energetic muons produced
in charged current interactions. The direction of these neutrinos can
only be reconstructed to $\sim 1.2^{\circ}/\sqrt{E_{\mu}(\rm{TeV})}$,
the angle at which the muon and neutrino are aligned. The angular cuts
which can be imposed when searching for a signal from the Sun are 
thus considerably weaker than for the liquid Argon technique described
in this article. Overcoming the atmospheric neutrino background 
is an essential requirement if a liquid Argon TPC detector is to succeed.

Experiments such as AMANDA, ANTARES and IceCube are (or will be) 
considerably larger than the 100 kton liquid Argon TPC detector we 
use here. The target mass of IceCube is approximately $10^4$ times
greater, for example. For very massive WIMPs (heavier than 100-200
GeV), high-energy neutrino telescopes provide a considerably more 
sensitive probe of dark matter annihilations than the technique 
described here. For lighter WIMPs, however, this may not be the case.

Considering a 50 GeV WIMP, for example, we expect neutrinos to be
generated with energies on the order of 15-20 GeV and lower (assuming 
the annihilation modes discussed in section~\ref{sec:WIMPsun}). Although such
neutrinos are well within the energy range that a liquid Argon
detector could be sensitive to, they would generate muons with
energies of 8-10 GeV, which are below the threshold for planned 
high-energy neutrino telescopes. Neutrinos from WIMPs lighter than 
$\sim 100$ GeV are likely to be missed by high-energy experiments, 
but are ideal for a liquid Argon TPC detector.

For discussions of the prospects for the detection of dark matter 
with high-energy neutrino telescopes, see 
Refs.~\cite{Bergstrom:1998xh,Barger:2001ur,Feng:2000zu,Bertin:2002ky}.

\section{Conclusions}
\label{sec:conclusion}

 In this paper we have studied the ability of a giant Liquid
Argon Time Projection Chamber (TPC) to detect dark matter indirectly via observation of neutrinos produced in WIMP annihilation in the core of the Sun. We have adopted an approach which differs from previous
studies, where an excess of upward-going muons would signal the
presence of WIMPs in the core of the Sun or the Earth. 
Our method takes advantage of the excellent angular reconstruction and 
superb electron identification capabilities Liquid Argon
offers to look for an excess of 
energetic electron (anti)neutrinos pointing in the direction of the
Sun. The expected signal and background event rates have been evaluated, in
a model independent way, as a function of the WIMP's elastic scatter cross
section for a range of masses up to 100 GeV.

 We have also presented a prospective study that quantifies the
detector discovery potential, i.e. the number of years needed to
claim a WIMP signal has been discovered. 
With the assumed set-up and thanks to the low background environment 
offered by the LAr TPC, a clear WIMP signal would be detected
provided the elastic scattering cross section in the Sun is above
$\sim 10^{-4}$~pb.

\bigskip

\acknowledgments

This work has been supported by the CICYT Grant FPA2002-01835.
S.N. acknowledges support from the Ramon y Cajal Program. 
D.H. is supported by the Leverhulme Trust. T.W. is 
supported by DOE grant DE-FG05-85ER40226. A.B. and T.W. 
thank the Benasque Center for Physics where the early stages 
of this work started.




\end{document}